\documentclass[epj,twocolumn]{webofc}
\usepackage[varg]{txfonts}   
\usepackage{hyperref}

\newcommand{\matriplex}{\textsc{Matriplex}\xspace}
\newcommand{\cplusplus}{{\small\textsc{C++}}\xspace}
\DeclareRobustCommand{\orderof}{\ensuremath{\mathcal{O}}}

\woctitle{Connecting The Dots 2016}
\begin{document}
\title{Kalman Filter Tracking on Parallel Architectures}

\author{\firstname{Giuseppe} \lastname{Cerati}\inst{1}\fnsep\thanks{\email{giuseppe.cerati@cern.ch}} \and
            \firstname{Peter} \lastname{Elmer}\inst{2}\fnsep\thanks{\email{peter.elmer@cern.ch}} \and
            \firstname{Slava} \lastname{Krutelyov}\inst{1}\fnsep\thanks{\email{vyacheslav.krutelyov@cern.ch}} \and
            \firstname{Steven} \lastname{Lantz}\inst{3}\fnsep\thanks{\email{steve.lantz@cornell.edu}} \and	
            \firstname{Matthieu} \lastname{Lefebvre}\inst{2}\fnsep\thanks{\email{ml15@princeton.edu}} \and	
            \firstname{Kevin} \lastname{McDermott}\inst{3}\fnsep\thanks{\email{kevin.mcdermott@cern.ch}} \and	
            \firstname{Daniel} \lastname{Riley}\inst{3}\fnsep\thanks{\email{daniel.riley@cornell.edu}} \and	
            \firstname{Matev\v{z}} \lastname{Tadel}\inst{1}\fnsep\thanks{\email{matevz.tadel@cern.ch}} \and	
            \firstname{Peter} \lastname{Wittich}\inst{3}\fnsep\thanks{\email{wittich@cornell.edu}} \and	
            \firstname{Frank} \lastname{W\"{u}rthwein}\inst{1}\fnsep\thanks{\email{fkw@ucsd.edu}} \and	
            \firstname{Avi} \lastname{Yagil}\inst{1}\fnsep\thanks{\email{ayagil@physics.ucsd.edu}}	
}

\institute{UC San Diego, 9500 Gilman Dr., La Jolla, California, USA 92093
\and
           Princeton University, Princeton, New Jersey, USA 08544
\and
           Cornell University, Ithaca, New York, USA 14850
          }

\abstract{%
Power density constraints are limiting the performance improvements of modern CPUs. To address this we have seen the introduction of lower-power, multi-core processors such as GPGPU, ARM and Intel MIC. To stay within the power density limits but still obtain Moore's Law performance/price gains, it will be necessary to parallelize algorithms to exploit larger numbers of lightweight cores and specialized functions like large vector units. Track finding and fitting is one of the most computationally challenging problems for event reconstruction in particle physics. At the High-Luminosity Large Hadron Collider (HL-LHC), for example, this will be by far the dominant problem. The need for greater parallelism has driven investigations of very different track finding techniques such as Cellular Automata or Hough Transforms. The most common track finding techniques in use today, however, are those based on the Kalman Filter. Significant experience has been accumulated with these techniques on real tracking detector systems, both in the trigger and offline. They are known to provide high physics performance, are robust, and are in use today at the LHC. We report on porting these algorithms to new parallel architectures. Our previous investigations showed that, using optimized data structures, track fitting with a Kalman Filter can achieve large speedups both with Intel Xeon and Xeon Phi. Additionally, we have previously shown first attempts at track building with some speedup. We report here our progress towards an end-to-end track reconstruction algorithm fully exploiting vectorization and parallelization techniques in a simplified experimental environment.
}

\maketitle
\section{Introduction}
\label{sec:intro}
The Large Hadron Collider (LHC) at CERN is the highest energy collider ever constructed. It consists of two counter-circulating proton beams made to interact in four locations around a 27 kilometer ring straddling the border between Switzerland and France. It is by some measures the largest man-made scientific device on the planet. The goal of the LHC is to probe the basic building blocks of matter and their interactions. In 2012, the Higgs boson was discovered by the CMS and ATLAS collaborations~\cite{cmshiggs,atlashiggs}. Experimentally, we collide proton beams at the center of our detectors and, by measuring the energy and momentum of the escaping particles, infer the existence of massive particles that were created and decayed in the pp collision and measure those massive particles' properties. In all cases, track reconstruction, i.e., the determination of the trajectories of charged particles (``tracks'') from a set of positions of energy deposits from the various layers in our detector (``hits''), plays a key role in identifying particles and measuring their charge and momentum.  Track reconstruction, also known as tracking, as a whole is the most computationally complex and time consuming, most sensitive to increased activity in the detector, and traditionally, least amenable to parallelized processing. The speed of online reconstruction has a direct impact on how much data can be stored from the 40 MHz collisions rate, while the speed on the offline reconstruction limits how much data can be processed for physics analyses. This research is aimed at vastly speeding up tracking.  

\begin{figure}[h]
\centering
\includegraphics[width=0.5\textwidth]{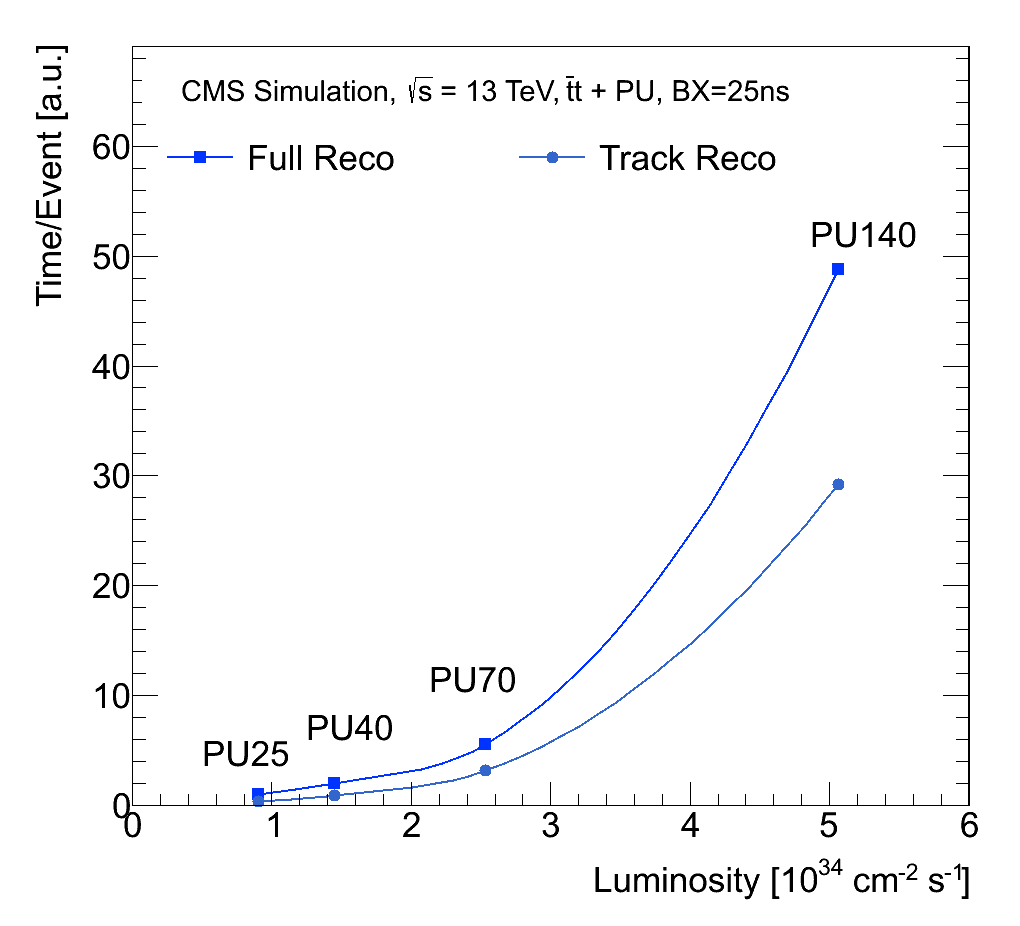}
\caption{CPU time per event versus instantaneous luminosity, for both full reconstruction and the dominant tracking portion. PU25 corresponds to the data taken during 2012, and PU140 corresponds to the HL-LHC era. The time of the reconstruction is dominated by track reconstruction. }
\label{fig:pileup}
\end{figure}

The large time spent in tracking will become even more important in the HL-LHC era of the Large Hadron Collider. The increase in event rate will lead to an increase in detector occupancy (``pile-up'', PU), leading to an exponential gain in time taken to perform track reconstruction, as can be seen in Fig.~\ref{fig:pileup}~\cite{vertex}. In the Figure, PU25 corresponds to the data taken during 2012, and PU140 corresponds to the low end of estimates for the HL-LHC era. Clearly this research on tracking performance will become increasingly important during this era.

A correlated issue to speeding up tracking is the change in computing architectures in the last decade. Around 2005, the computing processor market reached a turning point: power density limitations in chips ended the long-time trend of ever-increasing clock speeds, and our applications no longer immediately run exponentially faster on subsequent generations of processors. This is true even though the underlying transistor count continues to increase per Moore's law. Increases in processing speed such as those required by the time increases in Fig.~\ref{fig:pileup} will no longer come `for free' from faster processors. New, parallel processors instead are aggregates of `small cores' that in total still show large performance gains from generation to generation, but their usage requires a re-work of our mostly serial software to exploit these gains. The processors in question include ARM, GPGPU and the Intel Xeon and Xeon Phi; in this work we target the Xeon and Xeon Phi architectures.

\section{Kalman Filter Tracking}
\label{sec:kftrack}
The algorithm we are targeting for parallelized processing is a Kalman Filter (KF) based algorithm~\cite{Fruhwirth:1987fm}. KF-based tracking algorithms are widely used since they incorporate estimates of multiple scattering off massive detectors directly into the trajectory of the particle. Other algorithms, more naturally suited to parallelization and coming from the image processing community, have been explored by others. These include Hough Transforms and Cellular Automata, among others (see, for instance, ~\cite{HALYO1}.) However, these are not the main algorithms in use at the LHC today, whereas there is extensive understanding on how KF algorithms perform. KF algorithms have proven to be robust and perform well in the difficult experimental environment of the LHC~\cite{vertex}. Rather than abandon the collective knowledge gained from KF algorithms, we wish to extend this robust tool by porting KF tracking to parallel architectures. 

KF tracking proceeds in three main stages: seeding, building, and fitting. Seeding provides the initial estimate of the track parameters based on a few hits in a subset of the innermost layers. Realistic seeding is currently under development and will not be reported here. Building then collects additional hits in other detector layers to form a complete track, using the KF as a basis for predicting which hits to consider and keep. Track building is by far the most time consuming step of tracking, as it requires branching to explore potentially more than one candidate track per seed after finding compatible hits on a given layer. After hits have been assigned to tracks, a final fit using the KF is performed over each track to provide the best estimate of each tracks' parameters. 

KF tracking cannot be ported in a straightforward way to run in parallel on many-core processors for the following reasons. To realize performance gains, we need to exploit two types of parallelism: vectorization and parallelization. Vectorization aims to perform a single instruction on multiple data at the same time, in lockstep. In tracking, as explained above, we can branch to explore multiple candidates per seed, thus destroying the lock-step synchronization. Parallelization aims to perform different instructions at the same time on different data. The challenge to tracking then is workload balancing across different threads, as track occupancy in a detector is not uniformly distributed on a per event basis. Past work by our group has shown progress in porting sub-stages of KF tracking to support parallelism in simplified detectors (see, e.g. our presentations at ACAT2014~\cite{acat2014}, CHEP2015~\cite{chep2015}, and NSS-MIC2015~\cite{nss2015}).  As the hit collection is completely determined after track building, track fitting can repeatedly apply the KF algorithm without branching, making this the ideal place to start in porting KF tracking to Xeon and Xeon Phi, with our first results shown at ACAT2014~\cite{acat2014}. 

\subsection{Optimized Matrix Library \matriplex}
\label{sec:matriplex}
The computational problem of KF-based tracking consists of a sequence of matrix operations on matrices of sizes from $N\times{}N = 3\times{}3$ up to $N\times{}N = 6\times{}6$. To allow maximum flexibility for exploring SIMD operations on small-dimensional matrices, and to decouple the specific computations from the high level algorithm, we have developed a new matrix library, \matriplex. The \matriplex memory layout is optimized for the loading of vector units for SIMD operations on a set of matrices. \matriplex includes a code generator for defining optimized matrix operations, including support for symmetric matrices and on-the-fly matrix transposition. Patterns of elements which are known by construction to be zero or one can be specified, and the resulting generated code will be optimized accordingly to reduce unnecessary register loads and arithmetic operations. The generated code can be either standard \cplusplus or simple intrinsic macros that can be easily mapped to architecture-specific intrinsic functions.

\section{Track Building}
\label{sec:building}
Track building shares the same core KF calculations as track fitting, but has two major complications for utilizing vectorizable operations. The first such problem is the ``nHits'' problem: when performing the KF operations, the hit set for a given candidate track is undefined and one potentially has to choose between $\orderof{(10^4)}$ hits per layer. The second problem is the ``combinatorial'' problem: when more than one hit on a given layer is compatible with a given input candidate, branching occurs to explore all possible candidate tracks up to some cutoff number of candidates per seed. 

The key to reducing the nHits problem is to partition the tracks and hits spatially in order to reduce the search space for a given track. We partition space without reference to the detector structure, placing hits in each layer into directionally projective bins. We first define $\eta{}$ bins, where $\eta{} = -\log{[\tan^{-1}{(\theta{}/2)}]}$, $\theta{}$ being the polar angle with respect to the beam line. The $\eta{}$ bins are self-consistent, as in our detector, tracks do not bend in $\eta{}$. The bins of tracks are staggered with respect to bins of hits, so that tracks never search for hits outside two overlapping hit bins. This provides simple boundaries for thread associations using the OpenMP parallelization library~\cite{openmp}. We begin with 21 $\eta{}$ bins and distribute threads equally across the bins. We also define $\phi{}$ bins, where $\phi{}$ is the azimuthal angle, perpendicular to the beam line. As the hits are sorted in $\phi{}$ for every $\eta{}$ bin and layer, the $\phi{}$ bins simply provide a fast look-up of compatible hits in a layer for a given track.

With regards to the combinatorial problem, we first developed track building to only use the best hit out of all compatible hits on a layer for each candidate track. By definition, then, each seed only produces one track per layer and does not require copying of tracks to explore multiple hits per layer. The best hit is defined as the hit that produced the lowest $\chi{}^2$ increment to the track candidate. The vectorization and parallelization performance of this ``best hit'' approach were presented at CHEP2015~\cite{chep2015}. After demonstrating feasibility in the best hit approach, we then moved onto developing track building in the full combinatorial approach, which allows for exploring more than one track candidate per seed. 

\subsection{Memory Management}
\label{sec:memmgt}
With the full combinatorial approach in place, we performed extensive studies of the performance of our software, in terms of both the physics performance and the code performance. For the latter, we used the Intel VTune 2016~\cite{vtune} suite of tools to identify bottlenecks and understand the impacts of our optimization attempts. In particular, as can be expected, we determined that memory management is of critical importance.  To this effect we describe below several studies to optimize memory performance, and discuss the results of these studies.

The first such impact on performance arose out significant time being spent loading data into our local caches. By using vectorizable intrinsic functions to perform copying, we saw 20\% speedup. We also noticed a significant amount of time was being spent resizing the \textsc{C++} vector object that stored the indices of the hits associated to a given track, and simply reserving the memory did not help. By moving to a statically sized array, the resizing was avoided, yielding a 45\% speedup. The size of the data structures used in our algorithm has a crucial impact on the timing performance. In particular, the data structures for the hits and tracks must be optimized in size to fit into the fastest cache memory. We were able to reduce the size of the track objects by 20\% and the hit object by 40\%, leading to a 30\% speedup. There were additional objects used in the algorithm that were being instantiated after each event that could instead be reset and recycled, which led to a 25\% speedup. It is important to note that each of these improvements lead to individual, multiplicative speedups. 

Lastly, to mitigate the impact from serial work in the copying of track candidates in the combinatorial approach, we moved copying outside of all vectorizable operations in what we term the ``clone engine''. The clone engine approach only copies the best candidate track objects up to N candidates after reading and sorting a bookkeeping list of all compatible hits. This is in contrast to the original combinatorial approach which copied a candidate after each time a hit was deemed compatible, and then sorted and kept only the best N candidates after all the possible hits on a given layer for all input candidates were explored. A more detailed discussion of this work on memory management and impacts on performance was presented at NSS-MIC2015~\cite{nss2015}.

\subsection{Latest Results}
\label{sec:latest}	
\begin{figure*}
\centering
\includegraphics[width=1.0\textwidth]{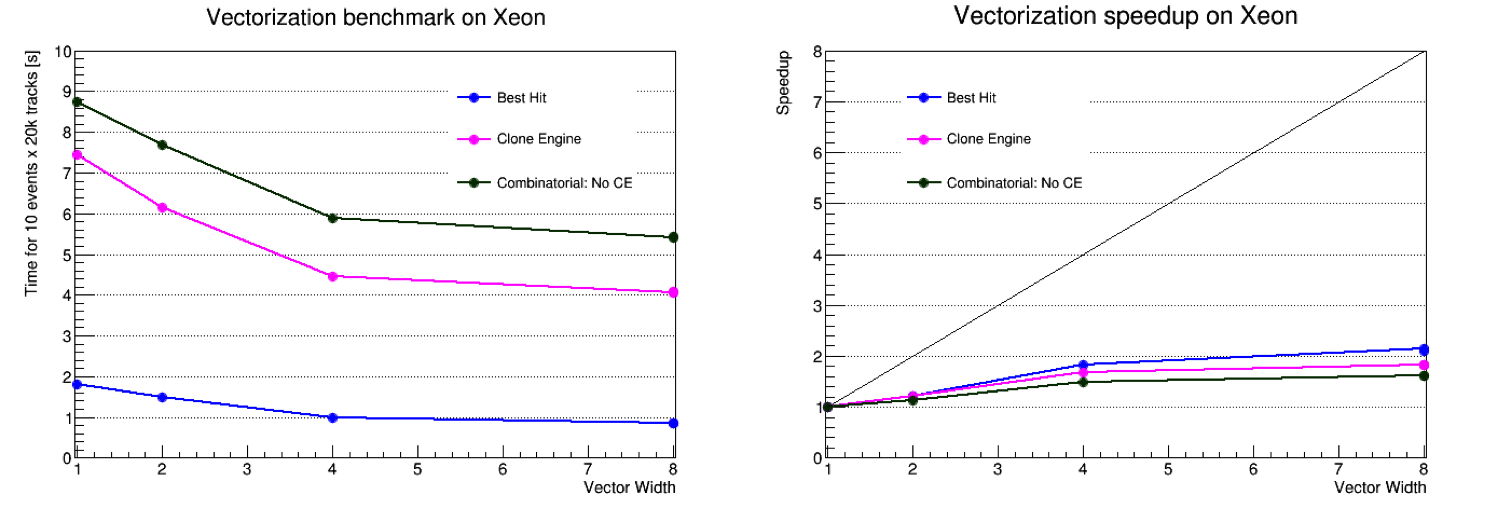}
\caption{Left: Absolute time as a function of the code parameter that sets the effective vector width. Right: Speedup as a function of this parameter on Xeon. Both plots show results for three different track building approaches. The blue curve is the no combinatorial tracking approach. The black and pink curves use the full combinatorial approach in track building. The pink curve moves the copying of tracks outside of the vectorizable operations. Only one thread is enabled.}
\label{fig:host_vu}       
\end{figure*}

We present here the latest vectorization and parallelization benchmarks in track building since NSS-MIC2015 given all of the developments described previously.   Figure~\ref{fig:host_vu} contains two plots displaying the vectorization performance of three different track building approaches on Xeon, shown as functions of the number of vector units enabled. The left plot displays the absolute time in seconds for building tracks for events with 20,000 tracks as a function of the code parameter that sets the effective width of the vector unit in floats. The plot on the right shows the speedup in time compared to the time from the first data point as a function of this parameter. Only one thread is enabled in these tests.

The blue curve is ``best hit'' approach described previously. Naturally, this approach will have the lowest absolute time in comparison to the two fully combinatorial approaches, the black and pink curves.  The black curve is the original approach to combinatorial track finding where the copying of the track candidates is inside the vectorizable KF operations. The pink curve shows the clone engine approach, which moves the serial work of copying outside of the KF vectorizable operations. As expected, the clone engine approach has a lower absolute time and higher speedup than the original. It is important to note that while the best hit approach is the fastest, the physics performance has inefficiencies in hit finding and track finding from not being fully able to explore multiple track candidates per seed. Even in our simplified model, this behavior is already apparent, and the best hit approach is expected to become even more inefficient as we add in more realistic detector effects.

\begin{figure*}
\centering
\includegraphics[width=1.0\textwidth]{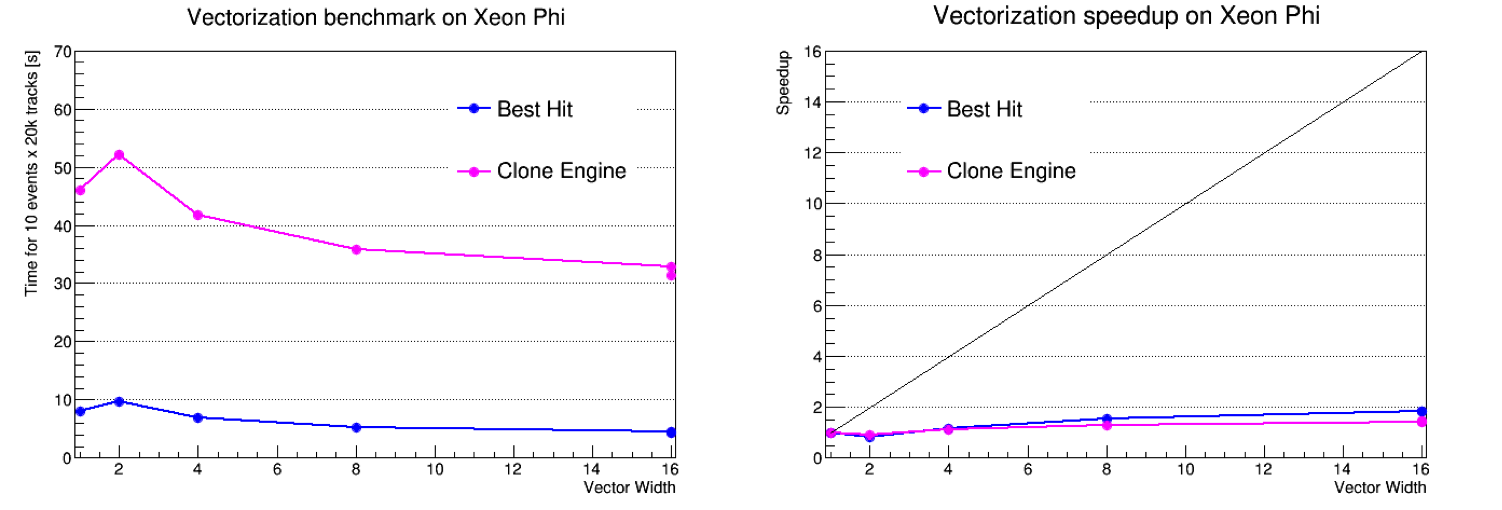}
\caption{Left: Absolute time as a function of the code parameter that sets the effective vector width on Xeon Phi. Right: Speedup as a function of this parameter on Xeon Phi. Both plots show results for two different track building approaches. The blue curve is the no combinatorial tracking approach. The pink curve uses the full combinatorial approach in track building and moves the copying of tracks outside of the vectorizable operations. Only one thread is enabled.}
\label{fig:mic_vu}       
\end{figure*}

Figure~\ref{fig:mic_vu} shows the same vectorization plots as Fig.~\ref{fig:host_vu}, now with Xeon Phi, which has AVX-512 vector registers that are twice as wide as those in Xeon. The original combinatorial approach is not shown, as it was apparent from Fig.~\ref{fig:host_vu} that it did not perform as well as the clone engine approach. With one thread enabled, Xeon Phi sees the same 2$\times{}$ speedup as Xeon. There is a noticeable increase in time after the vector unit size is first doubled. This is likely overhead from enabling the vector units. As seen on Xeon, the best hit approach still has the lowest absolute timing.

\begin{figure*}
\centering
\includegraphics[width=1.0\textwidth]{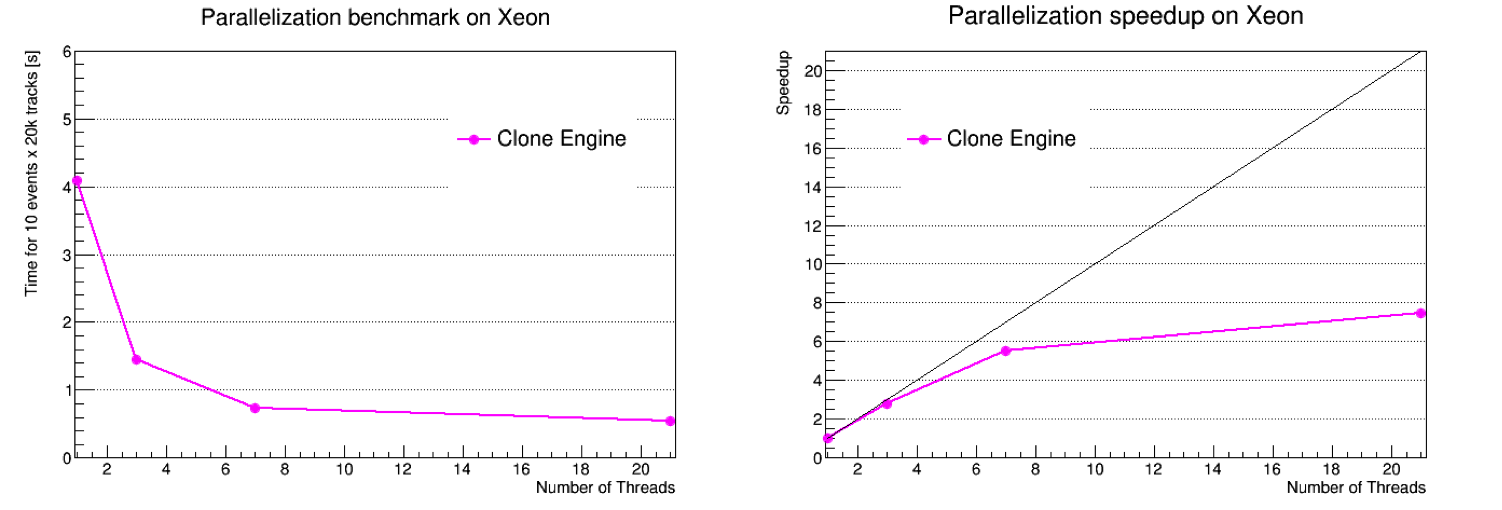}
\caption{Left: Absolute time as a function of the number of threads enabled on Xeon. Right: Speedup as a function of the number of threads enabled on Xeon. Both plots show results for the full combinatorial approach in track building that moves the copying of tracks outside of the vectorizable operations. The full Xeon vector width (8 floats) is assumed by the code.}
\label{fig:host_th}       
\end{figure*}

Figure~\ref{fig:host_th} displays the parallelization performance in terms of the absolute time and speedup of the clone engine approach on Xeon as a function of the number of threads enabled using OpenMP v. 4.0. The best hit approach is removed from these plots, because in reality, the combinatorial approach is one we will ultimately use. The Xeon machine we are using has 12 physical cores which appear to be 24 logical cores to the OS due to hyperthreading being enabled. As described previously, we enable threads based on the number of $\eta$ bins. We see near ideal speedup until we cross the boundary from physical to logical cores. The slight deviation away from ideal scaling even before  hyperthreading comes into play arises from the fact that some threads are unlucky and end up with more work per event than the rest, stalling the other threads, which was shown with VTune. The large deviation away from ideal scaling after enabling hyperthreading is due to resource limitations: the two threads per core are contending for the same instruction pipelines and data caches. Even so, nearly 8$\times{}$ speedup is seen for 21 threads.

\begin{figure*}
\centering
\includegraphics[width=1.0\textwidth]{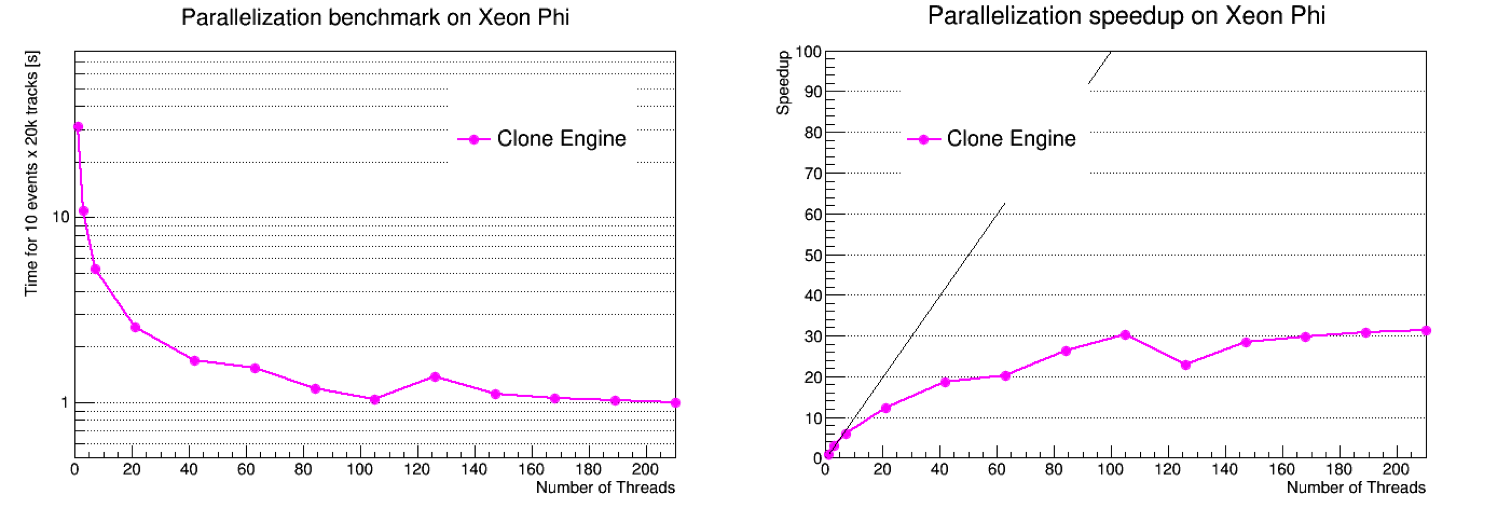}
\caption{Left: Absolute time as a function of the number of threads enabled on Xeon Phi. Right: Speedup as a function of the number of threads enabled on Xeon Phi. Both plots show results for the full combinatorial approach in track building that moves the copying of tracks outside of the vectorizable operations. The full Xeon Phi vector width (16 floats) is assumed by the code.} 
\label{fig:mic_th}       
\end{figure*}

Figure~\ref{fig:mic_th} displays the same parallelization performance plots as Fig.~\ref{fig:host_th}, now in terms of the Xeon Phi, with the full vector width of 16 floats enabled using OpenMP v. 4.0. The Xeon Phi we are using has 61 physical cores requiring 122 independent instruction streams for full utilization, due to the fact that the Xeon Phi issues instructions for a given thread every other clock cycle. Therefore, to keep a given physical core busy every clock cycle, one has to schedule two threads per core alternating in instruction execution. A form of hyperthreading is also present on Xeon Phi, yielding a total of 244 hardware threads (logical cores). Thus we notice two features in the speedup curve for Xeon Phi, occurring just past the physical core count (61) and twice that number (122). However, this loss in speedup is eventually recovered, and a factor of 30$\times{}$ speedup is observed with 210 threads enabled.

\section{Conclusion and Outlook}
\label{sec:conclusion}
We have made significant progress in parallelized and vectorized Kalman Filter-based end-to-end tracking R\&D on Xeon and Xeon Phi architectures. Through the use of a wide toolkit which includes Intel VTune, we have developed a good understanding of bottlenecks and limitations of our implementation which has led to further improvements. We are beginning to explore Intel Threaded Building Blocks (TBB), which allows for more dynamic scheduling to prevent stalls in resource usage. Additionally, threads can be spawned in numbers more naturally suited to the number of cores. Though it was not discussed in the talk, we have also developed tools to process fully realistic data, with encouraging preliminary results. Additionally, we have begun porting the \matriplex approach to GPGPU. The project has seen promising initial results; however, much work remains.  

\section{Acknowledgment}
\label{sec:ack}
This work is supported by the U.S. National Science Foundation, under the grants PHY-1520969, PHY-1521042, PHY-1520942 and PHY-1120138.

\end{document}